\documentclass[os]{copernicus}

\begin{document}

\title{Understanding mixing efficiency in the oceans: 
\\Do the nonlinearities of the equation
of state for seawater matter?}

\author[1]{R. Tailleux}

\affil[1]{Department of Meteorology, University of Reading, United Kingdom}


\runningtitle{Is mixing efficiency affected by nonlinear equation of state?}

\runningauthor{R. Tailleux}

\correspondence{R. Tailleux \\ (R.G.J.Tailleux@reading.ac.uk)}

\received{}
\pubdiscuss{} 
\revised{}
\accepted{}
\published{}


\firstpage{1}

\maketitle

\begin{abstract}
  There exist two central measures of turbulent mixing in
turbulent stratified fluids that are both caused by molecular diffusion:
1) the dissipation rate $D(APE)$ of available potential energy $APE$;
2) the turbulent rate of change $W_{r,turbulent}$ of background gravitational
potential energy $GPE_r$. So far, these two quantities have often been
regarded as the same energy conversion, namely the irreversible conversion
of $APE$ into $GPE_r$, owing to the well known exact equality $D(APE)=
W_{r,turbulent}$ for a Boussinesq fluid with a linear equation of state.
Recently, however, Tailleux (2009) pointed out that the above equality
no longer holds for a thermally-stratified compressible, with the ratio 
$\xi=W_{r,turbulent}/D(APE)$ being generally lower than unity
and sometimes even negative for water or seawater, and argued that $D(APE)$
and $W_{r,turbulent}$ actually represent two distinct types of energy
conversion, respectively the dissipation of $APE$ into one particular subcomponent
of internal energy called the `dead' internal energy $IE_0$, and the conversion
between $GPE_r$ and a different subcomponent of internal energy called
'exergy' $IE_{exergy}$. In this paper, the behaviour of the
ratio $\xi$ is examined 
for different stratifications having all the same buoyancy frequency
$N$ vertical profile, but different vertical profiles of the parameter
$\Upsilon = \alpha P/(\rho C_p)$,
where $\alpha$ is the thermal expansion coefficient, $P$ the
hydrostatic pressure, $\rho$ the density, and $C_p$ the specific heat capacity
at constant pressure, the equation
of state being that for seawater for different particular constant values of salinity.
It is found that $\xi$ and $W_{r,turbulent}$ depend critically on the sign
and magnitude of $d\Upsilon/dz$, in contrast with $D(APE)$, which appears largely
unaffected by the latter. These results have important consequences for how
the mixing efficiency should be defined and measured in practice, which are
discussed.
\end{abstract}


\introduction

  As is well known, turbulent diffusive mixing is a physical process that it is
  crucially important to parameterise well in numerical ocean models in order to
  achieve realistic simulations of the water mass properties and of the so-called
  meridional overturning circulation \citep{Gregg1987}, which are two essential
  components of the large-scale ocean circulation that may interact with Earth
  climate. For this reason, much effort has been devoted over the past decades
  toward understanding the physics of turbulent mixing in stratified fluids,
  one important goal being the design of physically-based parameterisations of 
  irreversible mixing processes for use in numerical ocean climate models.

  \par

  At a fundamental level, turbulent molecular diffusion in stratified fluids is 
  important for at least two distinct --- although inter-related --- reasons: 
  1) for transporting heat diffusively across isopycnal surfaces -- a process
  often referred to as `diapycnal mixing';
  2) for dissipating available potential energy, 
  which contributes for a significant
  fraction --- often called the mixing efficiency --- of the total dissipation of
  available mechanical energy $ME$, i.e., the sum of total kinetic energy
  $KE$ and available potential energy $APE$, which are defined by:
  \begin{equation}
        KE = \int_{V} \rho \frac{{\bf v}^2}{2} dV ,
  \end{equation}
  \begin{equation}
        APE = \underbrace{\int_{V} \rho \left ( g z + I \right ) dV}_{PE} - 
              \underbrace{\int_{V} \rho \left ( g z_r + I_r \right ) dV}_{PE_r} , 
  \end{equation} 
  where $\rho$ is the density, ${\bf v}=(u,v,w)$ is the three-dimensional velocity
  vector, $g$ is the acceleration of gravity, $z$ is the vertical coordinate 
  increasing upward, and $I$ the specific internal energy. The APE is defined as
  in \cite{Lorenz1955} as the difference between the potential energy $PE$ of the
  fluid (i.e., the sum of the gravitational potential energy $GPE$ plus 
  internal energy $IE$) minus the potential energy $PE_r$ of a reference state
  that is the state of minimum potential energy achievable in an adiabatic 
  re-arrangement of the fluid parcels. As shown by \cite{Winters1995}, the 
  $APE$ and $PE_r$ play a fundamental in the modern theory of turbulent mixing
  owing to the fact that by construction $PE_r$ is only affected by irreversible
  processes; as a result, measuring the time evolution of the reference state
  provides a direct and objective way to quantify the amount of irreversible 
  mixing taking place during turbulent mixing events, which is now commonly 
  exploited to diagnose mixing in numerical experiments, e.g., \cite{Peltier2003}.
  
  \par

 In the oceans, turbulent diapycnal mixing is required to transfer heat
 downward from the surface at a sufficiently rapid rate to balance the cooling of 
 the deep ocean by high-latitudes deep water formation. 
 In the oceanographic literature, the most widely used approach to parameterise
 the vertical (diapycnal) eddy diffusivity $K_{\rho}$ is based on the 
 Osborn-Cox model (\cite{Osborn1972}):
  \begin{equation}
        K_{\rho} = \frac{\varepsilon_P}{N^2} = 
        \frac{\gamma_{mixing} \,\,\varepsilon_K}{N^2} ,
        \label{Krho_osborn}
  \end{equation}
  which expresses $K_{\rho}$ in terms of either the turbulent
  viscous kinetic energy dissipation $\varepsilon_K$ or turbulent diffusive
  dissipation of available potential energy $\varepsilon_P$, where $N^2$ is
  the squared buoyancy frequency, and $\gamma_{mixing}=\varepsilon_P/
  \varepsilon_K$ is the ratio of the APE to KE dissipation, which is often
  called the `mixing efficiency', e.g. \cite{Lindborg2008}. Expressing 
  $K_{\rho}$ in terms of $\varepsilon_K$ appears to have been first proposed
  by \cite{Lilly1974} and \cite{Weinstock1978} in the context of stratospheric
  turbulent mixing, and adapted to the oceanographic case by \cite{Osborn1980}.
  The definition of mixing efficiency as a dissipations ratio adopted 
  in this paper appears to have been first proposed by 
  \cite{Oakey1982}.

  \par

  Since both $\epsilon_P$ and $\varepsilon_K$ are linked to the dissipation
  of mechanical energy of which $KE$ and $APE$ represent the two main 
  dynamically important forms,
  Eq. (\ref{Krho_osborn}) makes it clear that turbulent diapycnal mixing is
  directly related to the mechanical energy input in the oceans, but this
  link has been so far very rarely exploited in numerical ocean models.
  Rather, $K_{\rho}$ is often regarded as a tunable parameter whose value
  is adjusted to reproduce the main observed features of the oceanic
  stratification. Such an approach was used by 
  \cite{Munk1966}, who assumed the
  stratification to obey the vertical advective/diffusive balance:
  \begin{equation}
             w \frac{\partial \theta}{\partial z} = 
             \frac{\partial}{\partial z} 
             \left ( K_{\rho}\frac{\partial \theta}{\partial z}
             \right ),
             \label{advection_diffusion}
  \end{equation}
  where $\theta$ is the potential temperature, and $w$ the vertical velocity.
  Physically, Eq. (\ref{advection_diffusion}) states that the upward advection
  of cold water is balanced by the downward turbulent diffusion of heat, the
  rate of upwelling being set up by the rate of deep water formation.  
  By using Eq. (\ref{advection_diffusion}) as a model for stratification
  profiles in the Pacific,
  \cite{Munk1966} concluded that the canonical value
  $K_{\rho} = 10^{-4}\,{\rm m^2/s}$ was apparently needed to explain the
  observed structure of the oceanic thermocline. Subsequently,
  however, the validity of 
  \cite{Munk1966}'s approach was questioned, as several observational
  studies found $K_{\rho}$ in the ocean interior to be typically smaller by an
  order of magnitude than Munk's value, e.g., see \cite{Ledwell1998} and
  the review by \cite{Gregg1987}. However, it seems widely recognised today that
  $K_{\rho}$ is highly variable spatially, prompting 
  \cite{Munk1998} to re-interpret the value $K_{\rho}=10^{-4}\,{\rm m^2.s^{-1}}$
  as resulting from the overall effect of weak interior values combined with 
  intense turbulent mixing in coastal areas or over rough topography.
 
  \par

  While the above approach is useful, it does not exploit the link between 
  $K_{\rho}$ and the mechanical sources of stirring suggested by 
  Eq. (\ref{Krho_osborn}). Clarifying this link
  was pioneered by \cite{Munk1998}, who translated the advection/diffusion
  balance into one for the gravitational potential energy budget, 
   which they argue must be a balance between the rate of $GPE$ loss
    due to cooling and the rate of $GPE$ increase
    due to turbulent diffusive mixing, i.e., 
  \begin{equation}
           \left | \frac{d(GPE_r)}{dt} \right |_{cooling} \approx
           \left | \frac{d(GPE_r)}{dt} \right |_{mixing} , 
  \end{equation}
  this result being obtained by multiplying Eq. (\ref{advection_diffusion}) by
  $\alpha_{\theta} \rho_0 g z$, after some manipulation involving 
  integration by parts
  and the neglect of surface heating, where $\alpha_{\theta}$ is the thermal 
  expansion,
  $g$ the acceleration of gravity, $\rho_0$ a reference density, and $z$ the
  vertical coordinate pointing upward. The subscript $r$ is added here because
  it can be shown that \cite{Munk1998} must actually pertain to the background 
  $GPE_r$ budget, rather than the $AGPE$ budget, as shown by \cite{Tailleux2009}.
  This follows from the fact that cooling and turbulent molecular diffusion act
  as a $GPE$ sink and source only for the background $GPE_r$, as it is the 
  opposite that holds for $AGPE$. If one assumes that density is primarily
  controlled by temperature for simplicity, the effect of mixing on $GPE_r$ is
  thus given by:
  $$ 
     \left | \frac{d(GPE_r)}{dt} \right |_{mixing} = 
     - \int_{V} \rho_0 \alpha_{\theta} g z \frac{\partial}{\partial z}
       \left ( K_{\rho} \frac{\partial \theta}{\partial z} \right ) dV
  $$
  \begin{equation}
     = \int_{V} \rho_0 K_{\rho} N^2 \left ( 1 + \frac{z}{\alpha_{\theta}}
      \frac{\partial \alpha_{\theta}}{\partial z} \right ) dV , 
     \label{GPE_formula}
  \end{equation}
  by using the result that $N^2=\alpha_{\theta}g \partial \theta/\partial z$
  in absence of salinity effects, and by assuming $z=0$ at the ocean surface,
  and no flux through the ocean bottom.  In their paper, \cite{Munk1998}
  neglected the nonlinearities of the equation of state, 
   which amounts to regard $\alpha_{\theta}$ as constant, in which 
  case the above expression becomes:
  \begin{equation}
      \left | \frac{d(GPE_r)}{dt} \right |_{mixing} \approx
          \int_V \rho_0 K_{\rho} N^2 dV .
  \end{equation}
  By using Eq. (\ref{Krho_osborn}), assuming $\gamma_{mixing}$ 
  constant, this formula can be rewritten as follows:
  \begin{equation}
      \left | \frac{d(GPE_r)}{dt} \right |_{mixing}
     = \underbrace{\int_{V} \rho_0 \varepsilon_P dV}_{D(APE)} 
      = \gamma_{mixing} \underbrace{\int_{V} 
      \rho_0 \varepsilon_K dV}_{D(KE)}
      \label{closure}
  \end{equation}
  where $D(APE)$ and $D(KE)$ are the total volume-integrated diffusive dissipation
  of available potential energy and viscous dissipation of kinetic energy 
  respectively. To conclude, \cite{Munk1998} linked the dissipation to 
  production terms by assuming the balance $D(KE)=G(KE)$, where $G(KE)$
  is the work rate done by the mechanical forcing due to the winds and tides.
  As a result, the above formula yields:
  \begin{equation}
         G(KE) = \frac{1}{\gamma_{mixing}} \left |
         \frac{d(GPE)}{dt} \right |_{cooling} . 
        \label{GKE_constraint}
  \end{equation}
  By estimating the rate of $GPE$ loss due to cooling to be  
  $0.4\,{\rm TW}$, and by using the canonical value $\gamma_{mixing}=0.2$,
  \cite{Munk1998} concluded that 
  $G(KE)=O(2\,{\rm TW})$ of mechanical energy input was required to sustain 
  turbulent diapycnal mixing in the oceans. Since the work of the wind stress
  against the surface geostrophic velocity is widely agreed to be 
  $O(1\,{\rm TW})$, \cite{Munk1998} suggested that the shortfall should 
  be explained by the work rate done by the tides. The issue remains 
  controversial, however, because the role of the surface buoyancy forcing
  is not sufficiently well understood, as discussed in \cite{Tailleux2009}.

  \par

   Another important issue in assessing the uncertainties associated with
  Eq. (\ref{GKE_constraint}) concerns the importance of the nonlinearities
  of the equation of state, neglected by \cite{Munk1998}. As is well known,
  the nonlinearities of the equation of state are mostly responsible for 
  the fluid ``contracting upon mixing''. This contraction is responsible for
  the actual increase in $GPE_r$ due to mixing to be less than for a 
  linear equation
  of state. In Eq. (\ref{GPE_formula}), this can be seen from the fact that
  $\partial \alpha_{\theta}/\partial z$ is usually positive for a stably 
  stratified fluid. Since $z$ is negative by assumption, it follows that a
  correction factor is required that modifies Eq. (\ref{closure})
  as follows:
  $$
       \left | \frac{d(GPE)}{dt} \right |_{mixing} = (1-C_m) D(APE)
  $$
  \begin{equation}
      = (1-C_m) \gamma_{mixing} D(KE) .
  \end{equation}   
  where $C_m>0$, which in turn modifies \cite{Munk1998}'s constraint 
  (Eq. (\ref{GKE_constraint})) as follows:
  \begin{equation}
      G(KE) = \frac{1}{\xi \gamma_{mixing}} 
      \left | \frac{d(GPE)}{dt} \right |_{cooling}
  \end{equation}
  where $\xi = 1-C_m < 1$. Based on the above arguments, \cite{Munk1998}'s
  results are expected to underestimate the constraint on $G(KE)$, to the
  extent that
  $\gamma_{mixing}$ and the rate of $GPE_r$ loss due to cooling can be
  kept fixed. This point was first
  pointed out by \cite{Gnanadesikan2005} who emphasised the importance of
  cabelling. Discussing the value of $C_m$ or $\xi$ to be used in 
  Eq. (\ref{GKE_constraint}) is beyond the scope of this paper. Note, however,
  that it is possible to construct stratifications with $\xi$ not only smaller
  than one but also possibly even negative, as discussed by 
  \cite{Fofonoff1998,Fofonoff2001}. The latter cases are interesting, because
  they are such that $GPE_r$ {\em decreases} upon mixing, not {\em increases},
  in contrast to what is usually assumed.

  \par

  The main reason that $GPE_r$ is often assumed to increase as a result 
  of turbulent mixing stems from that for an incompressible fluid with a 
  linear equation of state, Eq. (\ref{closure}) states that:
  \begin{equation}
     \left | \frac{d(GPE)}{dt} \right |_{mixing} = D(APE), 
     \label{GPEr_DAPE}
  \end{equation}
  i.e., that $GPE_r$ increases at the same rate that $APE$ decreases, 
  which is classically interpreted as implying that the diffusively 
  dissipated $APE$ must be irreversibly converted into $GPE_r$, e.g., 
  \cite{Winters1995}. \cite{Tailleux2009} pointed out, however, that
  Eq. (\ref{GPEr_DAPE}) is at best only a good approximation, not a true
  equality, since in reality the rates of $GPE_r$ increase and $APE$ 
  decrease are never exactly equal, and sometimes even widely different,
  because of the nonlinear character of the equation of state.

  \par

  In order to better understand how the net change in $GPE_r$ correlates
  with the total amount of $APE$ diffusively dissipated during an 
  irreversible turbulent mixing event, it is useful to examine the 
  process of turbulent mixing 
  in the light of classical thermodynamic transformations.
  To make progress, the 
  conditions under which the diffusive exchange of heat between fluid parcels
  takes place need to be known, but in practice this is problematic, 
  for it would require solving the full compressible 
  non-hydrostatic Navier-Stokes equations down to the diffusive scales.
   Fortunately, it is often the case 
  that stratified fluids at low Mach numbers are close to hydrostatic
  equilibrium, suggesting that the diffusive heat exchange between 
  parcels may reasonably be assumed to occur at approximately constant
  pressure. If so, irreversible diffusive mixing must then be close to be a 
  process conserving the total potential energy $PE=APE+PE_r$ of the system,
  which implies that any amount $\Delta APE_{diff} < 0$ of diffusively 
  dissipated $APE$ must be irreversibly converted into background $PE_r$, viz.,
  \begin{equation}
      \Delta PE_r = -\Delta APE_{diff} > 0.
  \end{equation}
  The implications for the net change in $GPE_r$ can be determined 
  from the definitions  
  $PE_r = GPE_r + IE_r$, $APE = AGPE + AIE$, and $IE=AIE+IE_r$, which imply:
  \begin{equation}
       \Delta GPE_r = - \Delta APE_{diff} - \Delta IE_r
  \end{equation}
  where $\Delta IE_r$ is the net change in background
  internal energy taking place during
  the irreversible mixing event. As a result, the quantities $\xi$ and
  $C_m$ previously defined become:
  \begin{equation}
      \xi = \frac{\Delta GPE_r}{|\Delta APE_{diff}|} 
          = 1 - \frac{\Delta IE_r}{|\Delta APE_{diff}|} ,
      \label{xi_formula}
  \end{equation}
  \begin{equation}
      C_m = \frac{\Delta IE_r}{|\Delta APE_{diff}|} .
     \label{Cm_formula}
  \end{equation}
  Eqs. (\ref{xi_formula}) and (\ref{Cm_formula}) are important, because
  they establish that the nonlinearities of the equation of state --- which
  are responsible for the temperature and pressure dependence of $\alpha$ ---
  can give rise to internal energy changes $\Delta IE_r$
  comparable in magnitude with $\Delta APE_{diff}$ and $\Delta GPE_r$ 
  during a turbulent mixing event. Such large $IE_r$ changes must in turn
  be associated with potentially large compressibility effects whose work
  against the pressure field may also expected to be large, as first 
  demonstrated by \cite{Tailleux2009}. In other words, the above formula
  suggest that the nonlinearities of the equation of state may give rise
  to significant non-Boussinesq effects. So far, however, most numerical
  ocean models still make the incompressible and Boussinesq approximations,
  while at the same time using some version of the nonlinear equation of
  state for seawater. Such an approach yields values of $\xi$ and $C_m$
  that are predicted by Eq. (\ref{GPE_formula}), but since those values
  ultimately derive from initially making the Boussinesq approximation,
  it is unclear whether they can take into account the nonlinear character
  of the equation of state in a fully consistent manner.
  
  \par

  In fact, even when the net change $\Delta IE_r$ appears to be small
  or negligible, seemingly justifying the incompressible
  assumption, \cite{Tailleux2009} argues that compressible effects may
  still be large, because one may show that $\Delta IE_r$ can be  
  decomposed as follows:
  \begin{equation}
      \Delta IE_r = \Delta IE_{exergy} + \Delta IE_0,
  \end{equation}
  where $IE_0$ and $IE_{exergy}=IE_r-IE_0$ are two subcomponents of $IE_r$
  called the `dead' and `exergy' components. Physically, $IE_0$ represents
  the internal energy of a notional thermodynamic equilibrium state of uniform
  temperature $T_0$, whereas $IE_{exergy}$ represents the internal
  energy associated with the vertical stratification of the reference state.
  An important result of \cite{Tailleux2009} is that the net changes in 
  $IE_{exergy}$ and $IE_0$ are related at leading order to 
  $\Delta GPE_r$ and $\Delta APE_{diff}$ as follows: 
  \begin{equation}
      \Delta GPE_r \approx - \Delta IE_{exergy} ,
  \end{equation}
  \begin{equation}
      \Delta IE_0 \approx - \Delta APE_{diff}  > 0 ,
  \end{equation}
  to a very good approximation in a nearly incompressible fluid such 
  as water or seawater. These relations state that turbulent 
  molecular diffusion primarily dissipates $APE$ into `dead' internal 
  energy $IE_0$, while simultaneously causing a transfer between 
  $GPE_r$ and $IE_{exergy}$. Physically, the former effect results in an
  increase of the equivalent thermodynamic temperature $T_0$, whereas the
  latter effect results in the smoothing out of $dT_r/dz$. This contrasts
  with the standard interpretation that turbulent molecular diffusion
  irreversibly converts $APE$ into $GPE_r$, as proposed by \cite{Winters1995}.
  The differences between the two interpretations are schematically illustrated
  in Fig. 1. The main reason why compressibility effects may be important even
  if $\Delta IE_r \approx 0$ is because volume changes are primarily determined
  by $\Delta IE_{exergy}$, not by $\Delta IE_0$ or $\Delta IE_r$.  

  \par

  As regards to the empirical determination of the mixing efficiency
  $\gamma_{mixing} = \varepsilon_P/\varepsilon_K$, the above remarks are 
  important 
  because $\Delta GPE_r$ and $|\Delta APE_{diff}|$ are currently widely thought
  to physically represent the same quantity, prompting many studies to 
  actually estimate $\varepsilon_P$ from measuring the net changes in $GPE_r$,
  e.g., \cite{McEwan1983a,McEwan1983b,Barry2001}. For the reasons discussed 
  above, however, this makes sense only if $\xi$ can be ascertained to be
  close to unity, as if not, the relevant value of $\xi$ is then required.
  One of the main objective of this paper is to establish that the 
  behaviour of $\xi$ is closely connected to the sign and amplitude of the
  following parameter:
  \begin{equation}
         \frac{d}{dz} \left ( \frac{\alpha P}{\rho C_p} \right )
  \end{equation}
  where $\alpha$ is the thermal expansion coefficient, $P$ is the pressure,
  $\rho$ is the density, and $C_p$ is the specific heat capacity at constant
  pressure, while salinity is assumed to be uniform throughout the domain.
  Physically, the parameter $\Upsilon = \alpha P/(\rho C_p) 
  = P \Gamma/T$, where $\Gamma$ is the adiabatic lapse rate, represents the
  fraction of the amount of heat $\delta Q$ received by a parcel in an 
  isobaric process that can be converted into work. As a result, $\Upsilon$
  is expected to be the main parameter controlling the net change in $GPE_r$
  due to the turbulent diffusive heat exchange between fluid parcels.

  \par

  From the viewpoint of turbulent mixing, the main difficulty posed by a
  nonlinear equation of state is to make it possible for different 
  vertical stratification to share the same profile $N(z)$ without 
  necessarily having the same $\Upsilon(z)$ vertical profile. From a 
  dynamical viewpoint, this is not expected to be a problem as long as
  the dynamical evolution of $KE$ and $APE$, as well as $D(KE)$ and
  $D(APE)$, remain mostly controlled by $N(z)$ at leading order, as is
  usually assumed. If so, the dissipations ratio $D(APE)/D(KE)$, and hence
  the bulk mixing efficiency $\gamma_{mixing}$, can then be assumed to be
  unaffected by the nonlinearities of the equation of state at leading 
  order. The main objective of this paper is to verify that $D(APE)$ 
  appears indeed to be largely insensitive to the $\Upsilon(z)$ vertical profile,
  and hence mostly controlled by $N(z)$. If so, we can safely conclude
  that it must also be the case for $D(KE)$, since there is even less
  reasons to believe that the latter could be affected by $\Upsilon(z)$.
  This could be directly verified through direct numerical simulations
  of turbulent stratified mixing using a fully compressible Navier-Stokes
  equations solver, which we hope to report on in the future.
  On the other hand,
   the net change in $GPE_r$ is expected to be extremely sensitive to
  $\Upsilon (z)$. Most of the paper is devoted to verify that this is indeed
  the case, and to find ways to relate the net change in $GPE_r$ to the
  sign and magnitude of $d\Upsilon/dz$.
   Section 2 provides a theoretical formulation of the issue discussed.
  Section 3 discusses the methodology, while the results are presented in
  Section 4. Finally, section 5 summarises and discusses the results.

\section{Theoretical formulation of the problem}
     
  \subsection{Energetics of mixing}

     A key issue in the study of turbulent mixing is 
     understanding the links between stirring and mixing. As first discussed
     by \cite{Eckart1948}, the two processes can be rigourously separated if one    
     notes that the probability density functions (pdf in short) of the
     adiabatically conserved quantities (i.e., entropy and salt for seawater)
     are only affected by the irreversible mixing due to the molecular diffusion
     of heat and salt, but not by the adiabatic shuffling of the parcels 
     due to the stirring process. The link with \cite{Lorenz1955}'s
     available potential energy framework comes from the fact that Lorenz's 
     reference state, i.e., the state whose potential energy is minimised by
     an adiabatic re-arrangement of the fluid parcels, coincides with the 
     above-mentioned pdf, which was exploited by \cite{Winters1995} to provide
     a new way to rigourously quantify irreversible mixing simply from diagnosing
     the temporal evolution of the reference state. In a Boussinesq fluid with
     a linear equation of state, the role of entropy is played by either 
     temperature or density. In the following, the fluid will be assumed to
     be either freshwater or seawater with uniform salinity throughout the fluid.

     \par

       As shown by \cite{Winters1995} (using somewhat different notations),
     the energetics of freely decaying turbulence in an insulated domain is
     based on the following evolution equations for the volume-integrated
     kinetic energy $(KE)$, available potential energy $(APE)$, and background
     gravitational potential energy $(GPE_r)$:
     \begin{equation}
         \frac{d(KE)}{dt} = - C(KE,APE) - D(KE),
         \label{KE_evolution}
     \end{equation}
     \begin{equation}
         \frac{d(APE)}{dt} = C(KE,APE) - D(APE),
         \label{APE_evolution}
     \end{equation}
     \begin{equation}
         \frac{d(GPE_r)}{dt} = W_{r,mixing} = W_{r,laminar} 
          + W_{r,turbulent} ,
         \label{GPEr_evolution}
    \end{equation}
    where $C(KE,APE)$ is the so-called buoyancy flux, which physically
    represents the reversible conversion between $KE$ and $APE$, while
    all other terms represent irreversible processes, with $D(KE)$ 
    denoting the viscous dissipation of $KE$, $D(APE)$ the diffusive 
    dissipation of $APE$, and $W_{r,mixing}$ the rate of change of 
    $GPE_r$ due to molecular diffusion, which is customarily decomposed
    into a turbulent and laminar contribution. Note that the above equations
    are domain-averaged, not local formulations, which are expected to be
    well suited for understanding laboratory experiments of turbulent
    mixing for which lateral fluxes of $APE$ and $KE$ can be ignored.

    \par
 
    As discussed by \cite{Tailleux2009}, Eqs. (\ref{KE_evolution}-
    \ref{GPEr_evolution}) provide a unifying way to describe the energetics
    of both the incompressible Boussinesq and compressible Navier-Stokes
    equations, by adapting the definitions of the energy reservoirs and
    energy conversion terms to the particular set of equations considered.
    Explicit expressions for $D(APE)$ and $W_{r,mixing}$ are given by 
    \cite{Tailleux2009} in the particular cases of: 1) a Boussinesq fluid
    with a linear and nonlinear equation of state in temperature; 2)
    for a compressible thermally-stratified fluid obeying the Navier-Stokes
    equations of state with a general equation of state depending on 
    temperature and pressure. These expressions are recalled further below
    for case 2). While $W_{r,laminar}$ is well understood to be a conversion
    between $IE$ and $GPE_r$, the nature of the energy conversions associated with 
    $D(APE)$ and $W_{r,turbulent}$ is still a matter of debate. Currently,
    it is widely assumed that $D(APE)$ and $W_{r,turbulent}$ represent the same
    kind of energy conversion, namely the irreversible conversion of $APE$ into
    $GPE_r$ owing to the fact that for a Boussinesq fluid with a linear equation
    of state (referred to as the L-Boussinesq model hereafter), 
    one has the exact equality $D(APE)=W_{r,turbulent}$.
    It was pointed out by \cite{Tailleux2009} that this equality is a serendipitous
    artifact of the L-Boussinesq model, which does not hold for more accurate forms
    of the equations of motion. More generally, \cite{Tailleux2009} found that the
    ratio $\xi=W_{r,turbulent}/D(APE)$ is not only systematically lower than unity
    for water or seawater, but can in fact also takes on negative values, as 
    previously discussed by \cite{Fofonoff1962,Fofonoff1998,Fofonoff2001} in a series
    of little known papers. In other words, the equality $D(APE)=W_{r,turbulent}$
    is only a mathematical equality, not a physical equality, by defining 
    a physical equality as a mathematical equality between two quantities that
    persists for the most accurate forms of the governing equations of motion.
    To clarify the issue, \cite{Tailleux2009} sought to understand the links
    between $D(APE)$, $W_{r,mixing}$ and internal energy, 
    by establishing the following equations:
    \begin{equation}
        \frac{d(IE_0)}{dt} \approx D(KE) + D(APE),
        \label{IEo_evolution}
    \end{equation}
    \begin{equation}
        \frac{d(IE_{exergy})}{dt} \approx 
       = -\underbrace{[W_{r,laminar}  + W_{r,turbulent}]}_{W_{r,mixing}},
    \end{equation}
    which demonstrate that the viscously dissipated $KE$ and
    diffusively dissipated $APE$ both end up into the dead
    part of internal energy $IE_0$, whereas $W_{r,mixing}$
    represent the conversion rate between $GPE_r$ and the
    'exergy' component of internal energy $IE_{exergy}$.
    A schematic energy flowchart illustrating the above points
    is provided in Fig. 1.

    \subsection{Efficiency of mixing and mixing efficiency}

    The APE framework introduced by \cite{Winters1995} for a Boussinesq
    fluid with a linear equation of state, and extended by
    \cite{Tailleux2009} to a fully compressible thermally-stratified
    fluid, greatly simplifies the theoretical discussion of the concept
    of mixing efficiency. 
    To that end, it is useful to start with the evolution equation
    for the total ``available'' mechanical energy $ME=KE+APE$, 
    obtained by  summing the evolution equations for $KE$ and $APE$,
    leading to:
    \begin{equation}
       \frac{d(ME)}{dt} = -[D(KE)+D(APE)].
       \label{ME_evolution}
    \end{equation}
    Eq. (\ref{ME_evolution}), along with Eq. (\ref{IEo_evolution}), are
    very important, for they show that both viscous and diffusive 
    processes contribute to the dissipation of $ME$ into deal internal
    energy $IE_0$. From this viewpoint, understanding turbulent 
    diapycnal mixing amounts to understanding what controls the
    ratio $\gamma_{mixing}=D(APE)/D(KE)$, that is, the fraction of
    the total available mechanical energy dissipated by molecular
    diffusion rather than by molecular viscosity. The amount of ME
    dissipated by molecular diffusion, i.e., $D(APE)$, is important,
    because it is directly related to the definition of turbulent 
    diapycnal diffusivity, as said above in relation with 
    Eq. (\ref{Krho_osborn}).  

    \par   

    The link between the dissipation mixing efficiency and more 
    traditional definitions of mixing efficiency
    can be clarified in the light of the above energy
    equations, by investigating the energy budget of a notional ``turbulent
    mixing event'', defined here as an episode of intense mixing followed
    and preceded by laminar conditions (i.e., characterised by very weak
    mixing), during which $KE$ and $APE$ undergo a net change change 
    $\Delta KE<0$ and $\Delta APE<0$. As far as we understand the problem,
    most familiar definitions of mixing efficiency appear to implicitly
    assume $\Delta APE \approx 0$, as is the case for a turbulent mixing
    event developing from a unstable stratified shear flow for instance, 
    e.g., \cite{Peltier2003}. This point can be further clarified by 
    comparing the energetics of turbulent mixing events developing from
    the shear flow instability with that developing from the Rayleigh-Taylor
    instability, treated next, which by contrast can be regarded as having
    the idealised signature $\Delta KE \approx 0$ and $\Delta APE < 0$.

     \par 

     In the case of the stratified shear flow instability, assumed to 
    be such that $\Delta KE < 0$ and $\Delta APE \approx 0$, 
    integrating the above energy equations over the time interval over
    which the turbulent mixing event takes place 
    \footnote{It is usually assumed that the time average should be
    short enough that the viscous dissipation of the mean flow can be
    neglected. Alternatively, one should try to separate the laminar
    from the turbulent viscous dissipation rate. The following derivations
    assume that the viscous dissipation is dominated by the dissipation
    of the turbulent kinetic energy rather than that of the mean flow.} yields:
    \begin{equation}
       \Delta KE = - \overline{C(KE,APE)} - \overline{D(KE)},
       \label{ke_budget}
    \end{equation}
    \begin{equation}
       0 = \overline{C(KE,APE)} - \overline{D(APE)},
       \label{ape_budget}
    \end{equation}
    \begin{equation}
        \Delta GPE_r = \overline{W}_{r,mixing} = \overline{W}_{r,turbulent}
        + \overline{W}_{r,laminar},
   \end{equation}
   where the overbar denotes the time integral over the mixing event.
   For a Boussinesq fluid with a linear equation of state, \cite{Winters1995}
   showed that $\overline{D(APE)}=\overline{W}_{r,turbulent}$. If we combine
   the latter result with the $APE$ budget (i.e., Eq. (\ref{ape_budget})),
   one sees that one has the triple equality:
  \begin{equation}
      \overline{C(KE,APE)} = \overline{D(APE)} = \overline{W}_{r,turbulent}.
      \label{triple_equality}
  \end{equation}
  The triple equality Eq. (\ref{triple_equality}) suggests that any of the
  three quantities $\overline{C(KE,APE)}$, $\overline{D(APE)}$, or 
  $\overline{W}_{r,turbulent}$ can a priori serve to measure ``the fraction of
  the kinetic energy that appears as the potential energy of the stratification'', 
  which is the traditional definition of the flux Richardson number proposed by 
  \cite{Linden1979}. Historically, the buoyancy flux $\overline{C(KE,APE)}$ is
  the one that was initially regarded as the natural quantity to use for that
  purpose in an overwhelming majority of past studies of turbulent mixing.
  As a result, most existing studies of turbulent mixing define the turbulent
  diapycnal diffusivity, mixing efficiency, and flux Richardson number
  in terms of the buoyancy flux as follows:  
  \begin{equation}
         K_{\rho}^{flux} = \frac{\overline{C(KE,APE)}}{N^2}, 
         \label{krho_flux}
  \end{equation}
  \begin{equation}
        \gamma_{mixing}^{flux} = 
        \frac{\overline{C(KE,APE)}}{\overline{D(KE)}} ,
        \label{gamma_flux}
  \end{equation}
  \begin{equation}
         R_f^{flux} = \frac{\overline{C(KE,APE)}}{\overline{C(KE,APE)}+ 
         \overline{D(KE)}} .
  \end{equation}
  It is easily verified that the above equations are consistent with those
  considered by \cite{Osborn1980} for instance.
  Physically, however, there are fundamental problems
  in using the buoyancy flux to quantify irreversible diffusive mixing, because as
  pointed out by \cite{Caulfield2000}, \cite{Staquet2000} and \cite{Peltier2003},
  $C(KE,APE)$ represents a reversible energy conversion, which usually takes
  on both large positive and negative values before settling on its long term 
  average $\overline{D(APE)}$. Moreover, as pointed out below, the buoyancy
  flux is only related to irreversible diffusive mixing only if $\Delta APE\approx 0$
  holds to a good approximation, for otherwise, it becomes also related to 
  the irreversible viscous dissipation rate as shown by the KE budget 
  (Eq. (\ref{ke_budget})).  Eq. (\ref{triple_equality}) makes it
  possible, however, to use either $\overline{D(APE)}$ or 
  $\overline{W}_{r,turbulent}$ instead of $\overline{C(KE,APE)}$ in the
  definitions (\ref{krho_flux}) and (\ref{gamma_flux}). 
  For this reason, both \cite{Caulfield2000} and
  \cite{Staquet2000} proposed to measure the efficiency of mixing based on 
  $\overline{W}_{r,turbulent}$, i.e.,
  \begin{equation}
       K_{\rho}^{GPEr} = \frac{\overline{W}_{r,turbulent}}{N^2} ,
       \label{rho_gper}
  \end{equation}
  \begin{equation}
        \gamma_{mixing}^{GPEr} = \frac{\overline{W}_{r,turbulent}}{D(KE)} ,
        \label{gamma_gper}
  \end{equation}
  \begin{equation}
        R_f^{GPE_r} = \frac{\overline{W}_{r,turbulent}}
       {\overline{W}_{r,turbulent} + \overline{D(KE)}},
  \end{equation}
  such a definition being motivated by \cite{Winters1995}'s interpretation
  that $D(APE)$ and $W_{r,turbulent}$ represent the same energy conversion
  whereby the diffusively dissipated $APE$ is irreversibly converted into 
  $GPE_r$. The parameter $R_f^{GPE_r}$ was called the ``cumulative mixing
  efficiency'' by \cite{Peltier2003} and modified flux Richardson number
  by \cite{Staquet2000}. As argued in
  \cite{Tailleux2009}, it is $\overline{D(APE)}$, rather than 
  $\overline{W}_{r,turbulent}$, that directly measures the amount of $KE$ eventually
  dissipated by molecular diffusion via its conversion into $APE$, suggesting
  that the flux Richardson number should actually be defined as:
  \begin{equation}
       R_f^{DAPE}
       = \frac{\overline{D(APE)}}{\overline{D(KE)}+\overline{D(APE)}} .
       \label{me_def1}
  \end{equation}
  While the above formula makes it clear that all above definitions of $R_f$
  are equivalent in the particular case considered, it is easily realized 
  that they will in general yield different numbers if one relaxes the 
  assumption $\Delta APE \approx 0$ in Eq. (\ref{ape_budget}), as well as
  the assumption of a linear equation of state, yielding a ratio  
  $\xi = W_{r,turbulent}/D(APE)$ that is generally lower than unity and
  sometimes even negative for water or seawater. For this reason, it is 
  crucial to understand the physics of mixing efficiency at the most 
  fundamental level. From the literature, it seems clear that most 
  investigators's idea about the flux Richardson number is as a quantity
  comprised between $0$ and $1$. From that viewpoint,  
  the dissipation flux Richardson number $R_f^{DAPE}$ is the only 
  quantity that satisfies this property under the most general circumstances,
  as cases can easily be constructed for which 
  both $\overline{W}_{r,turbulent}$ and $\overline{C(KE,APE)}$ are negative.
  Indeed, cases for which $\xi<0$ are described in this paper, whereas 
  $\overline{C(KE,APE)}$ is easily shown to be negative in the case of
  a turbulent mixing event for which all mechanical energy is initially 
  provided entirely in $APE$ form. In that case, assuming $\Delta APE <0$
  and $\Delta KE \approx 0$ in the above energy budget equations yields:
  \begin{equation}
        \overline{C(KE,APE)} = \overline{D(APE)}+ \Delta APE = 
       - \overline{D(KE)} ,
  \end{equation}
  which shows that this time,
  $\overline{C(KE,APE)}$ directly measures the amount
  of viscously dissipated kinetic energy, rather than diapycnal mixing.
 The latter case is relevant to understand the energy budget of
 the Rayleigh-Taylor instability, see \cite{Dalziel2008} for a recent
 discussion of the latter.
 
  \subsection{Link between $D(APE)$ and $W_{r,mixing}$}

  In order to help the reader understand or appreciate why the 
  ratio $\xi=W_{r,turbulent}/D(APE)$ is generally lower than unity
  for water or seawater, and hence potentially significantly different
  from the predictions of the L-Boussinesq model, it is useful to 
  examine the structure of $W_{r,mixing}$ and $D(APE)$ in more
  details. As shown by \cite{Tailleux2009}, the analytical formula for 
  the latter quantities in a fully compressible
  thermally-stratified fluid are given by:
  \begin{equation}
          W_{r,mixing}= \int_{V} \frac{\alpha_r P_r}{\rho_r C_{pr}}
           \nabla \cdot \left (   \kappa \rho C_p \nabla T \right ) dV,
          \label{wrmixing_definition}  
\end{equation}
  \begin{equation}
       D(APE) = - \int_{V} \frac{T-T_r}{T} \nabla \cdot 
       \left ( \kappa \rho C_p \nabla T \right ) dV ,
       \label{dape_definition}
 \end{equation}
 where as before 
 $\alpha$ is the thermal expansion coefficient, $P$ is the pressure,
 $C_p$ is the specific heat capacity at constant pressure, $\rho$ is density,
 with the subscript $r$ indicating that values have to be estimated in their
 reference state. The parameter $\Upsilon = \alpha P/(\rho C_p)$ plays an
important role in the problem. Physically, it can be shown that in an isobaric
process during which the enthalpy of the fluid parcel increases by $dH$, 
the parameter $\Upsilon$ represents the fraction of $dH$ that is not converted
into internal energy, i.e., the fraction going into work (and hence contributing
ultimately to the overall net change in $GPE_r$). As a result, $\Upsilon$ 
plays the role of a Carnot-like thermodynamic efficiency.
In Eq. (\ref{wrmixing_definition}), $\Upsilon_r$ 
denotes the value that $\Upsilon$ would have if the corresponding fluid parcel
was displaced adiabatically to its reference position.

 \par
    In order to compare these two quantities, we expand $T$ as a
  Taylor series around $P=P_r$, viz.,
 \begin{equation}
     T = T_r + \Gamma_r (P-P_r) + \dots
 \end{equation}
 where $\Gamma_r = \alpha_r T_r/(\rho_r C_{pr})$ is the adiabatic lapse
 rate. At leading order, therefore, one may rewrite $D(APE)$ as follows:
$$
    D(APE) = \int_{V} \frac{\alpha_r (P_r-P)}{\rho_r C_{pr}}\frac{T_r}{T} 
         \nabla \cdot (\kappa \rho C_p \nabla T) dV + \dots
$$
$$
    = W_{r,mixing} + \int_{V} \frac{(T_r-T)}{T}\frac{\alpha_r P_r}{\rho_rC_{pr}}
    \nabla \cdot \left ( \kappa \rho C_p \nabla T \right ) dV 
$$
\begin{equation}
    - \int_{V} \frac{\alpha_r T_r P}{\rho_r C_{pr}T}
    \nabla \cdot \left ( \kappa \rho C_p \nabla T \right )\,dV  + \cdots
\end{equation}
These formula shows that $D(APE)$ can be written as the sum of $W_{r,mixing}$
plus some corrective terms. One sees that the L-Boussinesq model's results
derived by \cite{Winters1995} can be recovered in the limit
$T\approx T_r$, $P\approx -\rho_0 g z$, 
$\alpha_r/(\rho_r C_{pr})\approx \alpha_0 /(\rho_0 C_{p0})$, 
$\rho C_p \approx \rho_0 C_{p0}$, where the
subscript $0$ refers to a constant reference Boussinesq value, yielding:
\begin{equation}
  D(APE) \approx W_{r,mixing} - W_{r,laminar} = W_{r,turbulent} .
\end{equation}
These results, therefore, demonstrate that the strong correlation between
$D(APE)$ and $W_{r,mixing}$ originates in both terms depending on 
molecular diffusion in a related, but nevertheless distinct, way, the
differences between the two quantities being minimal for a linear equation
of state. The fact that the two terms are never exactly equal in a real fluid
clearly refutes \cite{Winters1995}'s widespread interpretation that $D(APE)$
and $W_{r,turbulent}$ physically represents the same energy conversion
whereby the diffusively dissipated $APE$ is irreversibly converted into $GPE_r$. 
In reality, $D(APE)$ and $W_{r,turbulent}$ represent two distinct types of 
energy conversions that happen to be both controlled by stirring and molecular
diffusion in related ways, which explains why they appear to be always strongly
correlated, and even exactly equal in the idealised limit of the L-Boussinesq
model. If one accepts the above point, then it should be clear that what is now
required to make progress is the understanding of what controls the behaviour
of the parameter $\xi$, since the knowledge of the latter is obviously crucial
to make inferences about turbulent diapycnal mixing from measuring the net 
changes of $GPE_r$ for instance. The purpose of the
numerical simulations described next is to help gaining insights into what 
controls $\xi$.

  \section{Methodology}
  
  To get insights into how the equation of state of seawater affects
  turbulent mixing, we compared $D(APE)$ and $W_{r,turbulent}$
  for a number of different stratifications having the same buoyancy
  frequency vertical profile $N$, but different vertical profiles with
  regard to the parameter $\alpha P/(\rho C_p)$, as illustrated in Fig. 2.
The quantities $D(APE)$ and $W_{r,mixing}$ were estimated from 
Eqs. (\ref{wrmixing_definition}) and (\ref{dape_definition}),
while $W_{r,turbulent}$ was estimated from
  \begin{equation}
          W_{r,turbulent} = W_{r,mixing} - W_{r,laminar} ,
  \end{equation}
where $W_{r,laminar}$ was obtained by taking $T=T_r$ in the
expression for $W_{r,mixing}$. The quantities $D(APE)$ and
$W_{r,turbulent}$ were estimated numerically for 
 a two-dimensional square domain discretised equally in the horizontal
 and vertical direction.
 In total, 27 different stratifications were considered, all possessing
 the same squared buoyancy frequency $N^2$ illustrated in the left panel
 of Figure 2, but different mean temperature, salinity, and pressure 
 resulting in different profiles for the $\alpha P/(\rho C_p)$ parameter
 illustrated in the right panel of Figure 2. In all cases considered, the
 pressure varied from $P_{min}$ to $P_{max}=P_{min} + 10 dbar$, 
 with $P_{min}$
 taking the three values (0 dbar, 1000 dbar, 2000 dbar). In all cases,
 the salinity was assumed to be constant, and taking one of the three
 possible values S = (30 Psu, 35 psu, 40 psu). With regard to the 
 temperature profile, it was determined by imposing the particular
 value $T_{max} = T(P_{min})$ at the top of the fluid, with
 all remaining values determined 
 by inversion of the buoyancy frequency $N^2$
 common to all profiles by an iterative method. The imposition of
 a fixed buoyancy profile $N$, salinity $S$, pressure range, and
 minimum temperature $T_{min}$ was found to yield widely
 different top-bottom temperature differences 
 $T(P_{min})-T(P_{max})$, ranging from a few tenths of degrees to about
 $4$ degrees C depending on the case considered, as seen in Fig. 3.
 In each case,
 the thermodynamic properties of the fluid were estimated from the
 Gibbs function of \cite{Feistel2003}. Specific details for the 
 temperature, pressure, and salinity in each of the 27 experiments
 can be found in Table 1 along with other key quantities discussed
 below.
 
 \par

 Numerically, the two-dimensional domain used to quantify 
 $D(APE)$ and $W_{r,turbulent}$ was discretised into 
 $N_{pi}\times N_{pj}$ points in the horizontal and vertical, with
 $N_{pi} = N_{pj} = 100$. Mass conserving coordinates were
 chosen in the vertical, and regular spatial Cartesian coordinate
 in the horizontal. For practical purposes, the vertical mass
 conserving coordinate can be regarded as standard height
 $z$, as the differences between the two types of coordinates
 were found to be insignificant in the present context, and thus
 chose $\Delta x = \Delta z$. In order to compute $D(APE)$ and
 $W_{r,turbulent}$ for turbulent conditions, we modelled the stirring
 process by randomly shuffling the fluid parcels adiabatically
 from resting initial conditions. Shuffling the parcels in such a way
 requires a certain amount of stirring energy, which is equal to
 the available potential energy $APE$ of the randomly shuffled
 state.
 
 \section{Results}

   For each of the 27 particular reference stratifications considered,
   synthetic turbulent states were constructed by generating hundreds of
   random permutations of the fluid parcels, thus simulating the effect
   of adiabatic shuffling by the stirring process, in each case yielding a
   particular value of $D(APE)$, $W_{r,mixing}$, $W_{r,turbulent}$ and 
   $APE$. One way to illustrate that $W_{r,turbulent}$ depends more 
   sensitively on the equation of state than $D(APE)$ is by plotting each
   quantity as a function of $APE$, as illustrated in Fig. 4. 
   Interestingly, the figure shows that all values of $D(APE)$ appear to
   be close to a linear straight line, with no obvious sensitivity to the
   particular value of $\Upsilon$. In contrast, the right panel of 
   Fig. 4 demonstrates the sensitivity of $W_{r,turbulent}$ to $\Upsilon$,
   as a separate curve is obtained for each different stratification.   
   Note that one should not construe from Fig. 4 that $D(APE)$ is a linear
   function of $APE$. Physically, $D(APE)$ depends both on the $APE$, as 
   well as on the spectrum of the temperature field. It so happens that 
   the method used to randomly shuffle the parcels tends to artificially
   concentrate all the power spectrum at the highest wavenumbers, the 
   effect of which being to suppress one degree of freedom to the problem,
   which is responsible for the appearance of a linear relationship 
   between $D(APE)$ and $APE$ in Fig. 4. It is easy to convince oneself,
   however, that stratifications can be constructed which have the same
   value of $APE$, but widely different values of $D(APE)$.

   \par

   In order to understand how the equation of state affects $W_{r,turbulent}$,
   it is useful to rewrite $W_{r,mixing}$ as given by 
   Eq. (\ref{wrmixing_definition}) as follows:
   $$
       W_{r,mixing} = - \int_{V} \kappa \rho C_p \nabla T \cdot
      \nabla \left ( \frac{\alpha_r P_r}{\rho_r C_{pr}} \right ) dV
   $$
   \begin{equation}
        \approx -\int_{V} \rho \kappa C_p  \frac{\partial}{\partial z_r} \left (
         \frac{\alpha_r P_r}{\rho_r C_{pr}} \right ) 
         \frac{\partial T_r}{\partial z_r} \| \nabla z_r \|^2 dV + \cdots 
         \label{wr_mixing}
   \end{equation}
   by using an integration by parts, assuming insulated boundaries,
   and using the approximation $\nabla T \approx 
   \nabla T_r + O(T-T_r)$, by noting that the reference quantities
   depend only upon $z_r$. Eq. (\ref{wr_mixing}) suggests that 
   $W_{r,mixing}$ and $W_{r,turbulent}$ are primarily controlled by 
   the vertical gradient of $\Upsilon = \alpha P/(\rho C_p)$, and that both
   $W_{r,mxing}$ and $W_{r,turbulent}$ are likely to be positive only when
   $d\Upsilon/dz$ is negative. This is obviously the 
   case when the vertical variations of $\alpha/(\rho C_p)$ can be neglected,
   as in this case $d\Upsilon/dz \approx \alpha/(\rho C_p) dP/dz
   \approx - \alpha g/C_p < 0$, assuming the pressure to be hydrostatic.
   The case when the vertical gradient of $\alpha P/(\rho C_p)$ is
   positive was extensively discussed by 
   \cite{Fofonoff1962,Fofonoff1998,Fofonoff2001}, and can be
   easily encountered in the oceans.

   \par

   In all experiments considered, we found the ratio $\xi=W_{r,turbulent}/D(APE)$
   to be systematically lower than unity, as already pointed out in 
   \cite{Tailleux2009}. In order to better understand
   how $d\Upsilon/dz$ controls the behaviour of $W_{r,turbulent}$, 
   the ratio $\xi = D(APE)/W_{r,turbulent}$ was averaged
   over all randomly shuffled states separately
   for each stratification, the results
   being summarised in Fig. 5 and Table 1, along with the minimum value of
   $d\Upsilon/dz$, as well as with the top-bottom difference $\Delta \Upsilon=
   \Upsilon(P_{min})-\Upsilon(P_{max})$.  Panels (a) and (c) show that as long
   that $d\Upsilon<0$, the equality $W_{r,turbulent}\approx D(APE)$ holds to
   a rather good approximation, up to a factor of 2, the approximation 
   being degraded at the lowest temperature and salinity. Note, however,
   that in the cases considered, $\xi>0$ only at atmospheric pressure, 
   with $\xi$ being systematically negative at $P_{min}=1000\,{\rm dbars}$
   and $P_{min}=2000\,{\rm dbar}$ respectively. Both Table 1 and 
   Fig. 5 (a) and (c) show that $\xi$ becomes increasingly negative as
   $[d\Upsilon/dz]_{min}$ becomes increasingly large and positive, 
   the worst case being achieved for the lowest $\overline{T}$, lowest
   salinity, and highest pressure. As a further attempt to understand
   this behaviour, we also computed the average ratio $AGPE/APE$ for
   each particular reference stratification. Interestingly, we find that
   the classical case $\xi \approx 1$ coincide with $APE\approx AGPE$,
   as expected in the Boussinesq approximation. We find, however, that 
   the decrease in $\xi$ coincides with $AGPE$ being an increasingly bad
   approximation of $APE$. As the latter implies that $AIE$ becomes
   increasingly important, it also implies that compressible effects 
   become increasingly important. This suggests, therefore, that 
   the effects of a nonlinear equation of state are apparently strongly
   connected to non-Boussinesq effects, a topic for future exploration. 
   
   \par
   
   The key point of the present results is that while there exist 
   stratifications such that $W_{r,turbulent}\approx D(APE)$ to a good
   approximation, and hence that conform to classical ideas about turbulent
   mixing in a Boussinesq fluid with a linear equation of state,
   there also exist stratification for which
   $W_{r,turbulent}$ and $D(APE)$ differ radically from each other. 
   The main reason why this is not more widely appreciated is suggested
   by the results summarised in Table 1, which shows that $W_{r,turbulent}
   \approx D(APE)$ appears to hold well under normal temperature and 
   pressure conditions, which are usually those encountered in most 
   laboratory experiments
   of turbulent mixing. In that case, the classical results
   of Boussinesq theory are applicable, and there is no problems in
   measuring the mixing efficiency of turbulent mixing events from measuring
   the net change in $GPE_r$, as often done, e.g., \cite{Barry2001}, in
   accordance with the definition of mixing efficiency proposed by 
   \cite{Caulfield2000} and \cite{Staquet2000}, since $\xi \approx 1$
   to a good approximation. Temperature, salinity, and pressure conditions
   in the real oceans can be very different than in the laboratory, however, 
   especially in the abyss. In the latter case, the present results suggest
   not only that $\xi$ can potentially become very large and negative, but
   that the discrepancy between $AGPE$ and $APE$ can become significant to
   the point of making the Boussinesq approximation and the neglect of 
   compressible effects very inaccurate. This point seems important in view
   of the current intense research effort devoted to understanding tidal
   mixing in the abyssal oceans that was prompted a decade ago by the 
   influential study by \cite{Munk1998}. The point is also important because
   values of mixing efficiency published in the literature have been 
   traditionally been reported without mentioning the associated value of
   $\xi$, which may explain part of the spread in the published values,
   and adds to the uncertainty surrounding this crucial parameter. 
   The present results suggest that an important project would be to 
   seek to reconstruct the missing values of $\xi$, which is in principle
   possible if sufficient information about the ambient conditions are
   available.

   \conclusions

   The nonlinearities of the equation of state for water or seawater
make it possible for 
a stratification with given mean vertical buoyancy profile $N$ to have
widely different vertical profiles of the parameter $\Upsilon = 
\alpha P/(\rho C_p)$, depending on particular oceanic circumstances.
The main result of this paper is that the sign
and magnitude of $d\Upsilon/dz$ greatly affect $W_{r,turbulent}$ ---
the turbulent rate of change of $GPE_r$ --- while they correspondingly 
little affect $D(APE)$, the dissipation rate of $APE$. As a result,
the ratio $\xi = W_{r,turbulent}/D(APE)$ is in general lower than unity,
and sometimes even negative, for water or seawater. For this reason, 
the fact that $D(APE)$ and $W_{r,turbulent}$ happen to be identical for a 
Boussinesq fluid with a linear equation of state appears to be a very
special case, which is rather misleading in that it fails to correctly
address the wide range of values assumed by the parameter $\xi$ 
in the actual oceans, while also leading to the widespread
erroneous idea that the diffusively dissipated $APE$ is irreversibly
converted into $GPE_r$, and hence that turbulent mixing always increase
$GPE$. As far as we understand the problem, based on the analysis
of \cite{Tailleux2009}, $D(APE)$ and $W_{r,turbulent}$ represent two
physically distinct kinds of energy conversion, the former associated
with the dissipation of $APE$ into `dead' internal energy, and the
latter associated with the conversion between $GPE_r$ and the 'exergy'
part of internal energy. The former is always positive, while the
latter can take on both signs, depending on the particular stratification.
   
   \par

   From the viewpoint of turbulence theory, the present results indicate
that the equality $D(APE)=W_{r,turbulent}$ obtained in the context of the
L-Boussinesq model by \cite{Winters1995} should only be construed as 
implying a strong correlation between $D(APE)$ and $W_{r,turbulent}$, not
as an indication that the diffusively dissipated $APE$ is converted into
$GPE_r$. As the present results show, the correlation between the two rates
strongly depends on the nonlinearities of the equation of state.
Fundamentally, $D(APE)$ and $W_{r,turbulent}$ appear to be correlated 
because they both depend on molecular diffusion, and on the gradient of
the adiabatic displacement $\zeta = z-z_r$ of the
isothermal surfaces from their reference positions. Based on the present
results, the ratio $\xi=W_{r,turbulent}/D(APE)$ appears to be determined
at leading order mostly by the sign and magnitude of $d\Upsilon/dz = d/dz[
\alpha P/(\rho C_p)$. Further work is required, however, to clarify the
precise link between $\xi$ and $d\Upsilon/dz$ under the most general 
circumstances, which will be reported in a subsequent paper.

   \par

   The present results are important, because they show that the two
following ways of defining a flux Richardson number $R_f$ and mixing
efficiency $\gamma_{mixing}$, viz.,
   \begin{equation}
          \gamma_{mixing}^{DAPE} = \frac{D(APE)}{D(KE)},
   \end{equation}
   \begin{equation}
          R_f^{DAPE} = \frac{D(APE)}{D(APE)+D(KE)}
   \end{equation}
   called the dissipation mixing efficiency and flux Richardson number by
   \cite{Tailleux2009}, and
   \begin{equation}
        \gamma_{mixing}^{GPEr} = \frac{W_{r,turbulent}}{D(KE)},
   \end{equation}
   \begin{equation}
         R_f^{GPEr} = \frac{W_{r,turbulent}}{W_{r,turbulent}+D(KE)},
   \end{equation}
   as proposed by \cite{Caulfield2000} and \cite{Staquet2000}, which are
equivalent in the context of the L-Boussinesq model, happen to be 
different in the context of a real compressible fluid, as the conversion
rules
   \begin{equation}
        \gamma_{mixing}^{GPEr} = \xi \gamma_{mixing}^{DAPE},
   \end{equation}
   \begin{equation}
        R_f^{GPEr} = \frac{\xi R_f^{DAPE}}{1-(1-\xi)R_f^{DAPE}} .
   \end{equation}
now involve the parameter $\xi$. Note that historically
the flux Richardson number
was defined by \cite{Linden1979} as ``The fraction of the kinetic energy
which appears as the potential energy of the stratification.'' Physically,
the kinetic energy that appears as the potential energy of the stratification
is the fraction of kinetic energy being converted into $APE$ and ultimately
dissipated by molecular diffusion. This fraction is therefore measured by
$D(APE)$, not by $W_{r,turbulent}$, since the latter technically represents
the ``mechanically-controlled'' fraction of internal energy converted into
$GPE_r$, if one accepts \cite{Tailleux2009}'s conclusions. From this 
viewpoint, it is $R_f^{DAPE}$ rather than $R_f^{GPE_r}$ that appears
to be consistent with \cite{Linden1979}'s definition of the flux Richardson
number, and hence $\gamma_{mixing}^{DAPE}$ rather than 
$\gamma_{mixing}^{GPE_r}$ that is consistent with \cite{Osborn1980}'s 
definition of mixing efficiency.

   \par

  From a practical viewpoint, however, the above conceptual objections
against $\gamma_{mixing}^{GPE_r}$ and $R_f^{GPE_r}$ do not mean that it
is equally physically objectionable to seek estimating the efficiency of
mixing from measuring the net changes in $GPE_r$ taking place during a
turbulent mixing event, as is commonly done, e.g., \cite{Barry2001}.
Such a method is perfectly valid, owing to the correlation between
$D(APE)$ and $W_{r,turbulent}$. The present results show, however, that
such an approach requires the knowledge of the parameter $\xi$, which 
is usually not supplied. For most laboratory experiments performed at
atmospheric pressure, the issue is probably unimportant, as $\xi$ 
appears to be generally close to unity in that case. The issue becomes
more problematic, however, for measurements carried out in the ocean
interior, as there is less reason to assume that $\xi \approx 1$ will
be necessarily verified. A critical review of published values of 
$\gamma_{mixing}$ would be of interest, in order to identify the cases
potentially affected by a value of $\xi$ significantly different from unity.

\par
   So far, we have only considered the case of an equation of state
   depending on temperature and pressure only, by holding salinity constant.
   In practice, however, many studies of turbulent mixing are based on
   the use of compositionally stratified fluids. Understanding whether
   $\xi$ can be significantly different from unity in that case remains
   a topic for future study.




\begin{acknowledgements}
The author acknowledges funding from the RAPID programme.
Rainer Feistel, Michael McIntyre and an anonymous referee are also thanked
for provided comments that led to significant improvements in
presentation and clarity.

\end{acknowledgements}




f
\begin{figure*}[t]
\vspace*{2mm}
\begin{center}
\includegraphics[width=15cm]{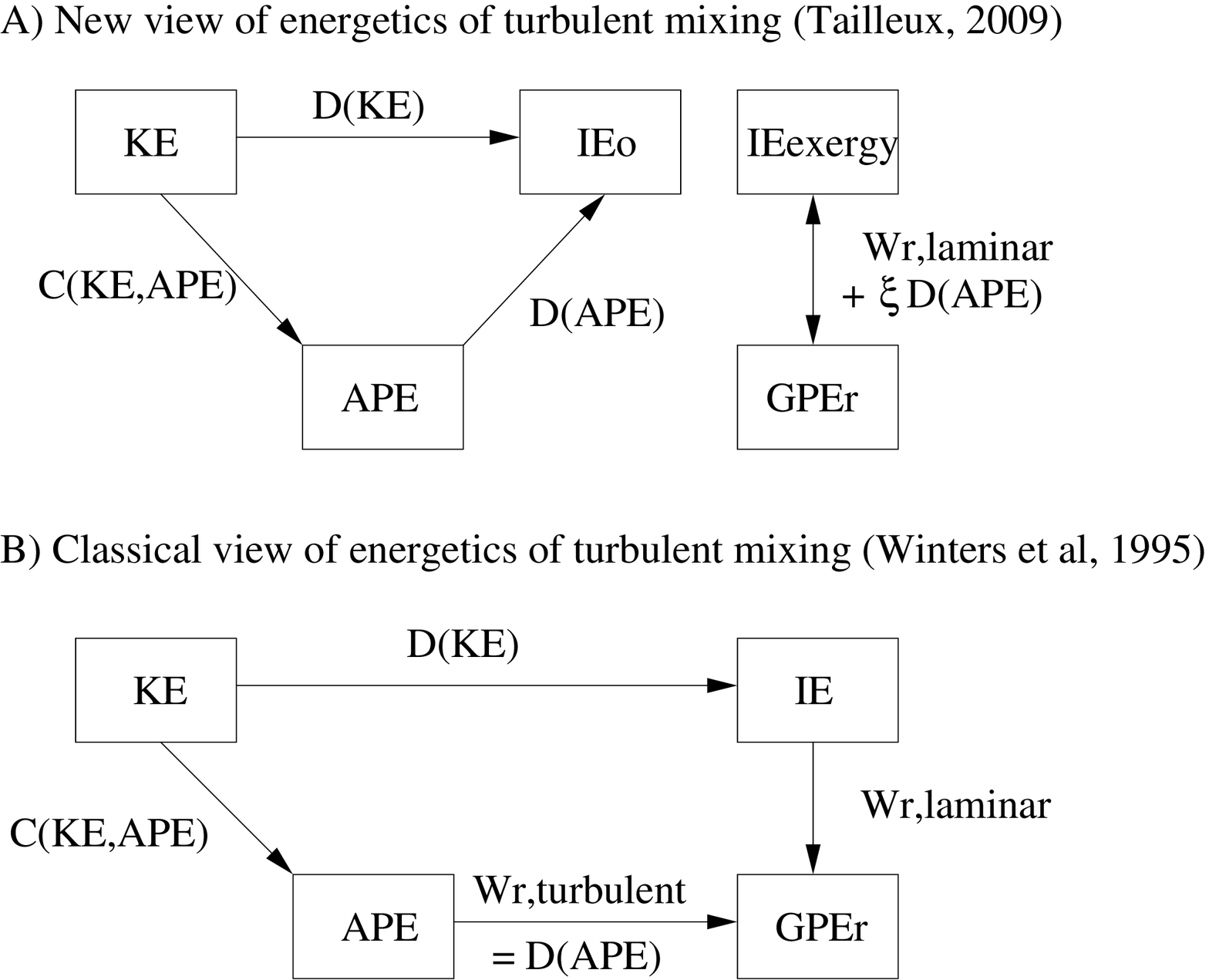}
\end{center}
\label{figure0}
\caption{A) New view of the energetics of freely decaying
turbulent stratified mixing as proposed by Tailleux (2009)
versus B) the earlier interpretation proposed by Winters et al. (1995).
In the new view, internal energy $IE$ is subdivided into a dead part
$IE_0$ and exergy part $IE_{exergy}$. The double arrow linking 
$IE_{exergy}$ and $GPE_r$ means that both $W_{r,laminar}$ and
$W_{r,turbulent}$ can be either positive or negative in general.}
\end{figure*}

f
\begin{figure*}[t]
\vspace*{2mm}
\begin{center}
\includegraphics[width=15cm]{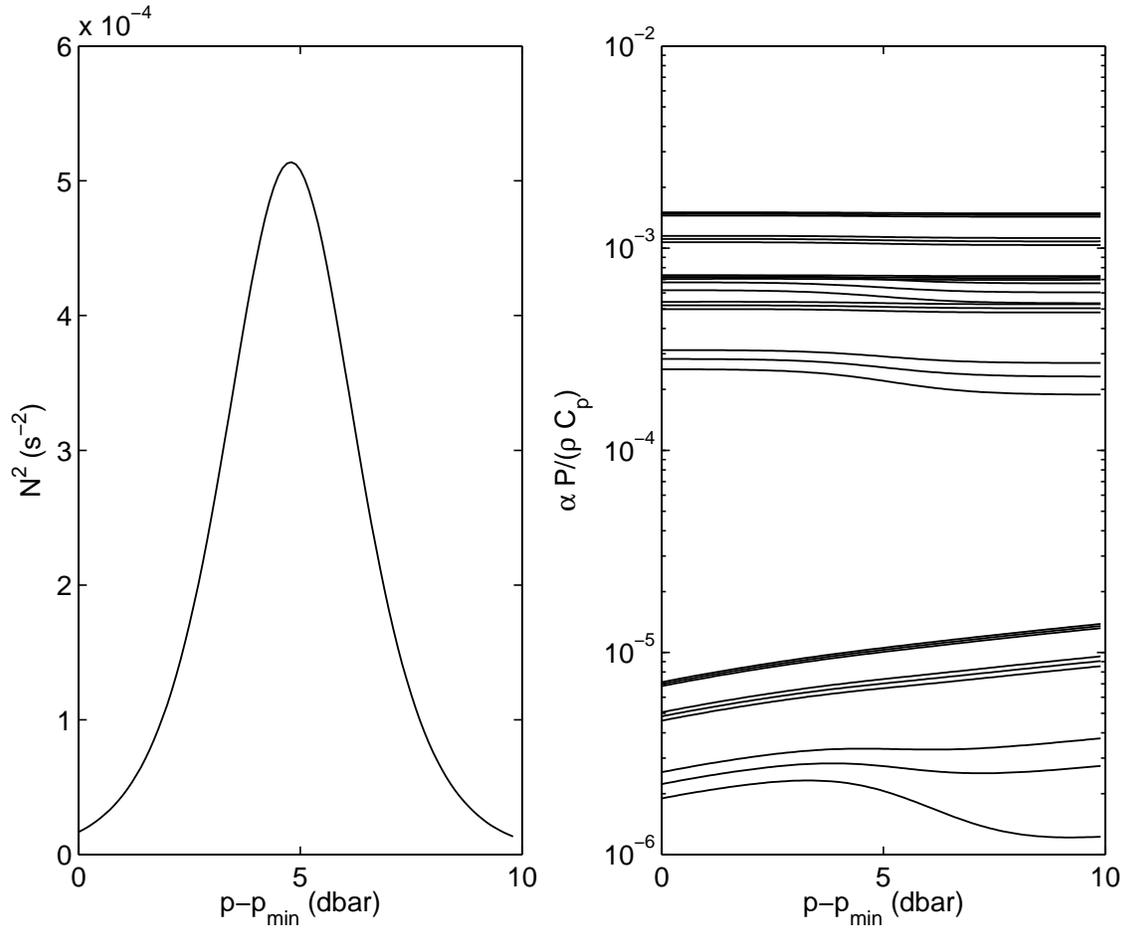}
\end{center}
\label{figure1}
\caption{(Top panel) The squared buoyancy frequency $N^2$ common to all
stratifications considered. (Bottom panel) The thermodynamic efficiency-like
quantity $\alpha P/(\rho C_p)$ corresponding to the 27 different cases 
considered. Note that the Fofonoff regime, i.e., the case for which $GPE$
decreases as the result of mixing, is expected whenever the latter quantity
decreases for increasing pressure. The classical case considered by the 
literature, i.e., the case for which $GPE$ increases as the result of mixing 
corresponds to the case where the latter quantity increases with increasing
pressure on average (see Table 1 for more details).}
\end{figure*}

\begin{figure*}[t]
\vspace*{2mm}
\begin{center}
\includegraphics[width=15cm]{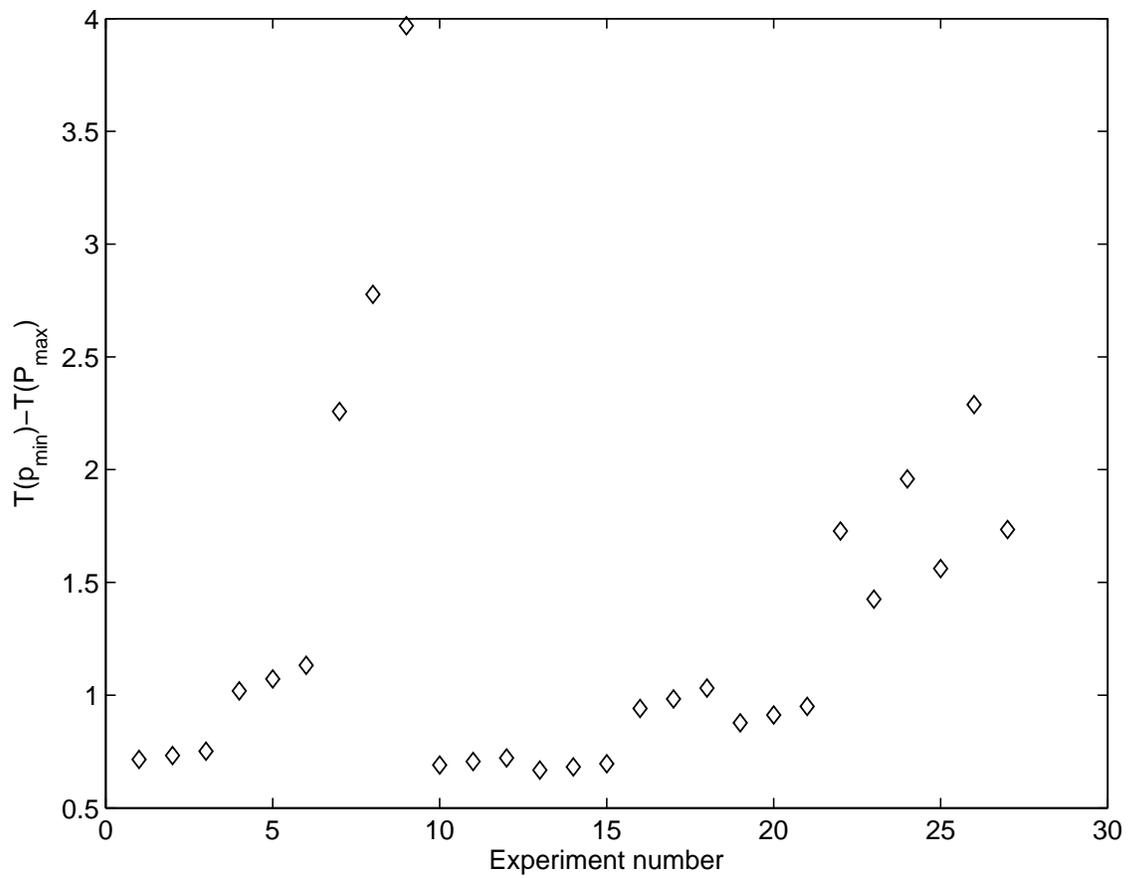}
\end{center}
\caption{Distribution of the top-bottom temperature difference
$T(P_{min})-T(P_{max})$ as a function of the experiment number.}
\end{figure*}


f
\begin{figure*}[t]
\vspace*{2mm}
\begin{center}
\includegraphics[width=15cm]{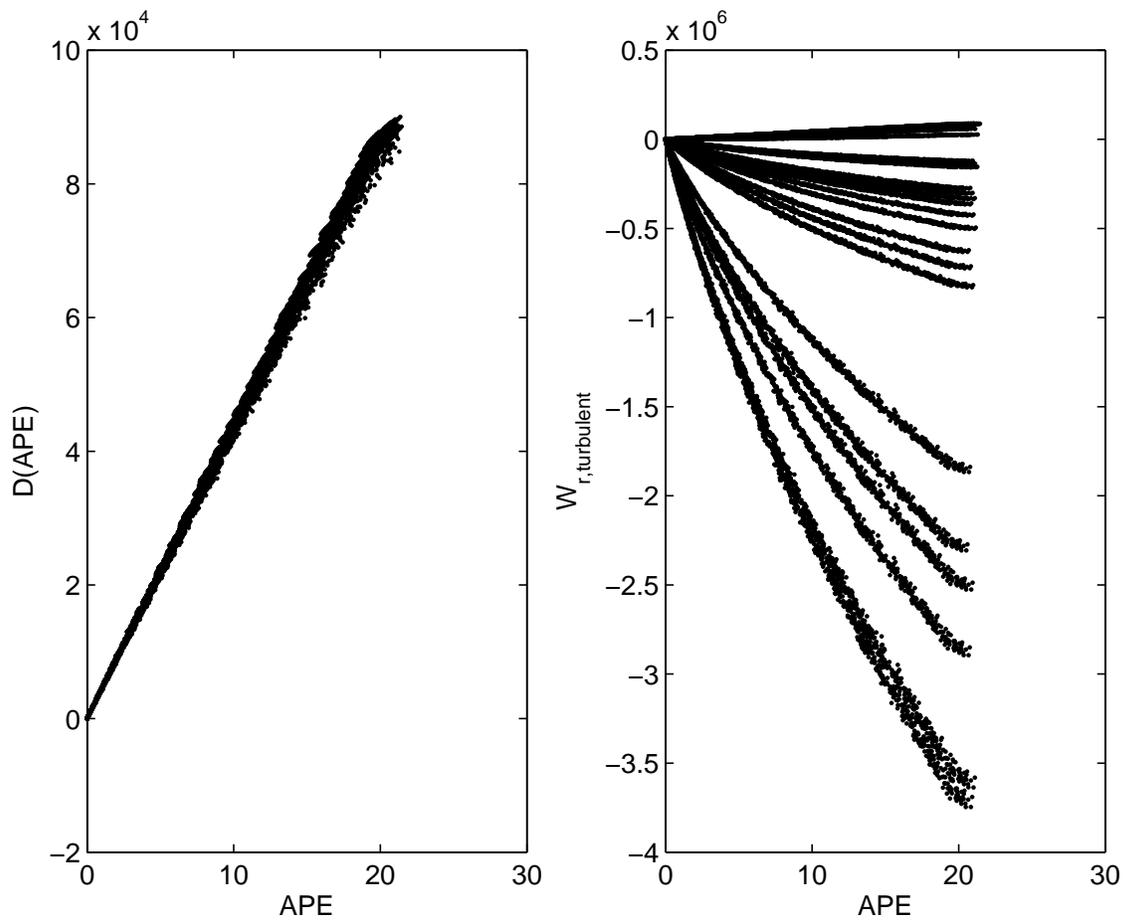}
\end{center}
\label{figure2}
\caption{(Left panel) The dissipation rate of $APE$ as a function of $APE$,
each point corresponding to one particular experiment. Note that there is
no obvious dependence on the stratification.
(Bottom panel) The rate of change $W_{r,turbulent}$ as a function of $APE$.
This time, each stratification is associated with a different curve.}

\end{figure*}

\begin{figure*}[t]
\vspace*{2mm}
\begin{center}
\includegraphics[width=15cm]{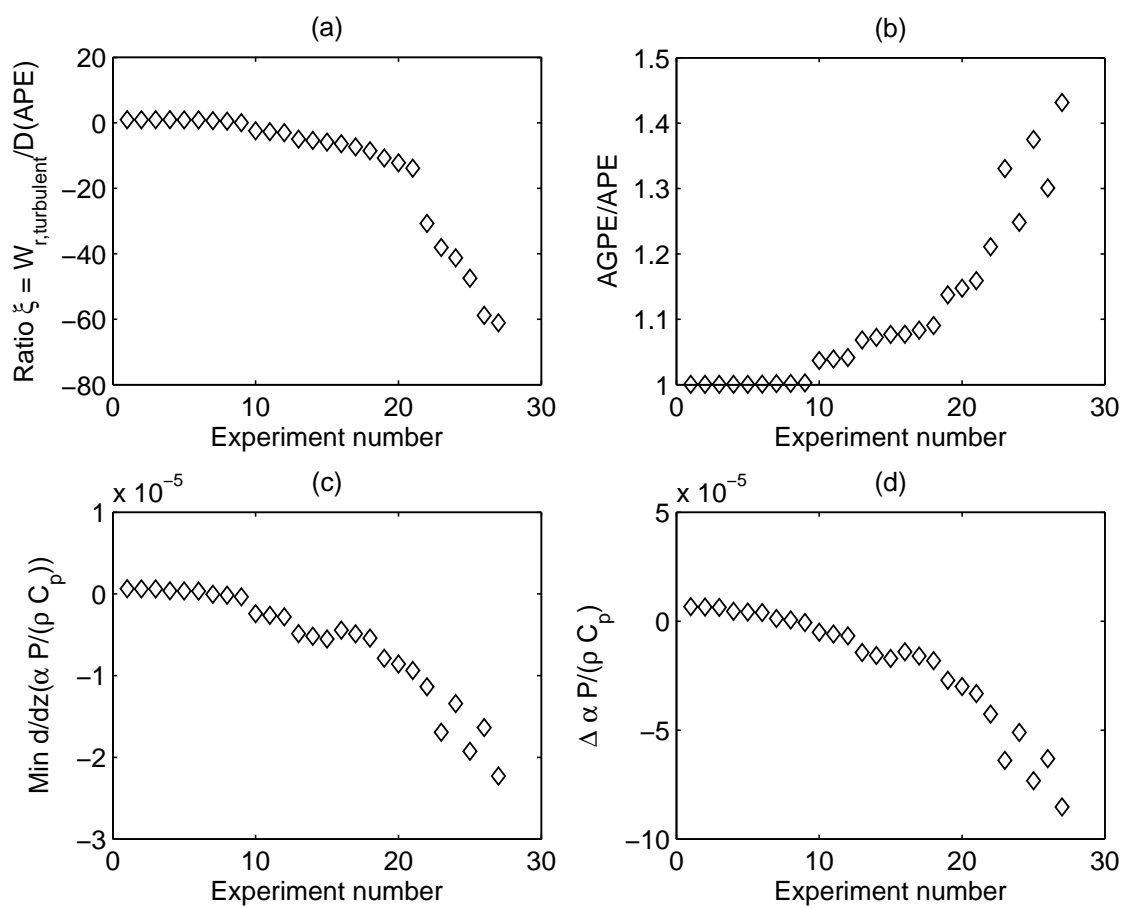}
\end{center}
\label{figure3}
\caption{(a) The averaged ratio $\xi=W_{r,turbulent}/D(APE)$ as a function of
the experiment number; (b) The averaged ratio $AGPE/APE$ as a function of
the experiment number; (c) The minimum value of $d/dz[\alpha P/(\rho C_p)$
as a function of the experiment number; (d) The top-bottom difference
of $\alpha P/(\rho C_p)$ as a function of the experiment number. 
}
\end{figure*}







t
\begin{table*}[t]
\caption{Averaged values of the two ratios $\xi=W_{r,turbulent}/D(APE)$ and
$AGPE/APE$ for the 27 different types of stratifications considered in this
paper. The quantities $[d\Upsilon/dz]_{min}$ and $\Delta \Upsilon$ refer to
the minimum value of the vertical derivative of $\Upsilon=\alpha P/(\rho C_p)$
and top-bottom difference of $\Upsilon$ respectively. $S$ is the salinity
used in the equation of state for seawater, $\overline{T}$ is the mean
temperature of the profile considered, and $P_{min}$ denotes the minimum
value of the vertical pressure profile. The top-bottom temperature
differences are displayed in Fig. 3, while the pressure interval
is $10\,{\rm dbar}$ in all cases. The tabulated values demonstrate that 
increasingly negative values of $\xi$ coincide with increasingly
large positive values of $d\Upsilon/dz$, as well as with
with the increasing importance of non-Boussinesq compressible 
effects associated with an increasing discrepancy between $AGPE$ and $APE$.
The standard case for which $\xi \approx 1$ is achieved close to atmospheric
pressure. The maximum negative value of $\xi$ occurs for the lowest 
$S$, lowest $\overline{T}$, and largest $P_{min}$ values considered.}
\vskip4mm
\centering
\begin{tabular}{lcccccrr}
\tophline
Expt & $\xi$ & $AGPE/APE$ & $[d\Upsilon/dz]_{min} \times 10^6$ 
  & $\Delta \Upsilon\times 10^6$ & 
  S(psu) &  $\overline{T}(^{\circ}C)$ & $P_{min}(dbar)$ \\
\middlehline
  1 &  0.98 & 1.0003 &  -6.70 &  -0.64 & 40 & 22.6 &   0 \\
  2 &  0.98 & 1.0003 &  -6.53 &  -0.63 & 35 & 22.6 &   0 \\ 
  3 &  0.98 & 1.0003 &  -6.36 &  -0.61 & 30 & 22.6 &   0 \\ 
  4 &  0.95 & 1.0005 &  -4.50 &  -0.40 & 40 & 12.5 &   0 \\ 
  5 &  0.95 & 1.0005 &  -4.23 &  -0.37 & 35 & 12.5 &   0 \\ 
  6 &  0.94 & 1.0006 &  -3.95 &  -0.33 & 30 & 12.4 &   0 \\ 
  7 &  0.71 & 1.0015 &  -1.20 & 0.03 & 40 &  1.9 &   0 \\ 
  8 &  0.55 & 1.0018 &  -0.51 & 0.15 & 35 &  1.6 &   0 \\
  9 &  0.10 & 1.0026 & 0.67 & 0.35 & 30 &  1.2 &   0 \\ 
 10 & -2.41 & 1.0369 & 5.07 & 2.42 & 40 & 22.6 & 1000 \\ 
 11 & -2.67 & 1.0391 & 5.89 & 2.61 & 35 & 22.6 & 1000 \\
 12 & -2.96 & 1.0416 & 6.76 & 2.81 & 30 & 22.6 & 1000 \\
 13 & -4.93 & 1.0682 & 14.42 &  4.87 &  40 & 22.7 & 2000 \\ 
 14 & -5.36 & 1.0724 & 15.72 &  5.18 &  35 & 22.7 & 2000 \\ 
 15 &  -5.84 & 1.0768 & 17.09 &  5.51 & 30 & 22.6 & 2000 \\ 
 16 &  -6.35 & 1.0772 & 14.05 &  4.44 & 40 & 12.5 & 1000 \\ 
 17 &  -7.35 & 1.0835 & 15.97 &  4.89 & 35 & 12.5 & 1000 \\ 
 18 &  -8.53 & 1.0905 & 18.10 &  5.40 &  30 & 12.5 & 1000 \\ 
 19 & -10.73 & 1.1372 & 27.10 &  7.87 &  40 & 12.6 & 2000 \\ 
 20 & -12.17 & 1.1476 & 30.02 &  8.58 &  35 & 12.5 & 2000 \\ 
 21 & -13.86 & 1.1591 & 33.23 &  9.37 &  30 & 12.5 & 2000 \\ 
 22 & -30.73 & 1.2109 & 42.67 & 11.36 &  40 &  2.1 & 1000 \\ 
 23 & -38.06 & 1.3306 & 63.86 & 16.93 &  40 &  2.3 & 2000 \\ 
 24 & -41.26 & 1.2482 & 51.06 & 13.42 &  35 &  2.0 & 1000 \\
 25 & -47.46 & 1.3751 & 73.20 & 19.26 &  35 &  2.2 & 2000 \\ 
 26 & -58.84 & 1.3010 & 63.09 & 16.37 &  30 &  1.9 & 1000 \\ 
 27 & -61.06 & 1.4318 & 85.31 & 22.28 &  30 &  2.1 & 2000 \\ 
\bottomhline
\end{tabular}
\label{only_table}
\end{table*}




\addtocounter{figure}{-1}\renewcommand{\thefigure}{\arabic{figure}a}


\begin{thebibliography}{}




   \bibitem[Barry (2001)]{Barry2001}
      \textsc{Barry, M.E., Ivey, G. N., Winters, K. B., and
       Imberger, J.} 2001
       Measurements of diapycnal diffusivities in stratified fluids.
       \emph{J. Fluid Mech.} \textbf{442}, 267--291.
   \bibitem[Caulfield \& Peltier (2000)]{Caulfield2000}
      \textsc{Caulfield, C.P. \& Peltier, W.R.} 2000
      The anatomy of the mixing transition in homogeneous and stratified
      free shear layers. \emph{J. Fluid Mech.}, \textbf{413}, 1--47.
   \bibitem[Dalziel \& al (2008)]{Dalziel2008}
\textsc{Dalziel, S. B., Patterson, M. D., Caulfield, C. P., \&
       Coomaraswamy, I. A.} 2008
       Mixing efficiency in high-aspect-ratio Rayleigh-Taylor experiments.
       \emph{Phys. Fluids}, \textbf{20}, 065106.
   \bibitem[Eckart (1948)]{Eckart1948}
      \textsc{Eckart, C.} 1948
      An analysis of the stirring and mixing processes in incompressible
      fluids. \emph{J. Mar. Res.}, \textbf{7}, 265--275.
   \bibitem[Feistel (2003)]{Feistel2003}
      \textsc{Feistel, R.} 2003
      A new extended Gibbs thermodynamic potential of seawater.
      \emph{Prog. Oceanogr.} \textbf{58}, 43--114.
   \bibitem[Fofonoff (1962)]{Fofonoff1962}
      \textsc{Fofonoff, N.P.} 1962
      Physical properties of seawater.
      M.N. Hill, Ed., \emph{The Sea}, Vol. 1, Wiley-Interscience, 3--30.
   \bibitem[Fofonoff (1998)]{Fofonoff1998}
      \textsc{Fofonoff, N.P.} 1998
      Nonlinear limits to ocean thermal structure. 
      \emph{J. Mar. Res.} \textbf{56}, 793--811.
   \bibitem[Fofonoff (2001)]{Fofonoff2001}
      \textsc{Fofonoff, N.P.} 2001
      Thermal stability of the world ocean thermoclines.
      \emph{J. Phys. Oceanogr.} \textbf{31}, 2169--2177.
   \bibitem[Gnanadesikan \& al. (2005)]{Gnanadesikan2005}
       \textsc{Gnanadesikan, A., Slater, R.D., Swathi, P.S., \&
       Vallis, G.K.} 2005
       The energetics of the ocean heat transport.
       \emph{J. Climate}, \textbf{18}, 2604--2616.
   \bibitem[Gregg (1987)]{Gregg1987}
       \textsc{Gregg, M. C.} 1987
       Diapycnal mixing in the thermocline: a review. 
       \emph{J. Geophys. Res.} \textbf{92-C5}, 5249--5286.
    \bibitem[Ledwell \& al (1998)]{Ledwell1998}
        \textsc{Ledwell, J. R., Watson, A. J. \& Law, C.S.} 1998
        Mixing of a tracer in the pycnocline.
        \emph{J. Geophys. Res. Oceans} \textbf{103}, 21499--21519.
    \bibitem[Lilly \& al. (1974)]{Lilly1974}
        \textsc{Lilly, D. K., Waco, D. E., \& Adelfang, S. I.} 1974
        Stratospheric mixing estimated from high-altitude turbulence
        measurements. \emph{J. Appl. Meteor.}, \textbf{13}, 488-493.  
   \bibitem[Lindborg \& Brethouwer (2008)]{Lindborg2008}
       \textsc{Lindborg, E. \& Brethouwer, G.} 2008
       Vertical dispersion by stratified turbulence.
       \emph{J. Fluid Mech.} \textbf{614}, 303--314.
   \bibitem[Linden (1979)]{Linden1979}
        \textsc{Linden, P. F.} 1979
        Mixing in stratified fluids.  
        \emph{Geophys. Astrophys. Fluid Dyn.} \textbf{13}, 3--23.
   \bibitem[Lorenz (1955)]{Lorenz1955}
        \textsc{Lorenz, E. N.} 1955
        Available potential energy and the maintenance of the general
        circulation. \emph{Tellus} \textbf{7}, 157--167.
     \bibitem[McEwan (1983a)]{McEwan1983a}
       \textsc{McEwan, A. D.} 1983a
       The kinematics of stratified mixing through internal wavebreaking.
       \emph{J. Fluid Mech.} \textbf{128} 47--57.
   \bibitem[McEwan (1983b)]{McEwan1983b}
       \textsc{McEwan, A. D.} 1983b
        Internal mixing in stratified fluids.
        \emph{J. Fluid Mech.} \textbf{128} 59--80.
   \bibitem[Munk (1966)]{Munk1966}
        \textsc{Munk, W.} 1966
        Abyssal recipes
        \emph{Deep-Sea Res.} \textbf{13}, 207-230. 
   \bibitem[Munk \& Wunsch (1998)]{Munk1998}
        \textsc{Munk, W. \& Wunsch, C.} 1998
        Abyssal recipes II: energetics of tidal and wind mixing.
        \emph{Deep-Sea Res.} \textbf{45}, 1977--2010.
 \bibitem[Oakey (1982)]{Oakey1982}
       \textsc{Oakey, N. S.} 1982
        Determination of the rate of dissipation of turbulent energy from
        simultaneous temperature and velocity shear microstructure measurements. 
       \emph{J. Phys. Oceanogr.} \textbf{12}, 256--217.
 \bibitem[Osborn and Cox (1972)]{Osborn1972}
      \textsc{Osborn, T. R. and Cox, C. S.} 1972
       Oceanic fine structure. \emph{Geophys. Astr. Fluid Dyn.}
       \textbf{3}, 321--345.   
\bibitem[Osborn (1980)]{Osborn1980}
        \textsc{Osborn, T. R.} 1980
        Estimates of the local rate of vertical diffusion from 
        dissipation measurements. \emph{J. Phys. Oceanogr.}
        \textbf{10}, 83--89.
   \bibitem[Paparella \& Young (2002)]{Paparella2002}
       \textsc{Paparella, F. \& Young, W. R.} 2004
       Horizontal convection is non turbulent.
       \emph{J. Fluid Mech.} \textbf{466}, 205-214.
   \bibitem[Peltier \& Caulfield (2003)]{Peltier2003}
        \textsc{Peltier, W. R. \& Caulfield, C. P.} 2003
         Mixing efficiency in stratified shear flows.
       \emph{Annu. Rev. Fluid Mech.} \textbf{35}, 135--167.
   \bibitem[Staquet (2000)]{Staquet2000}
        \textsc{Staquet, C.} 2000
       Mixing in a stably stratified shear layer: two- and 
       three-dimensional numerical experiments. \emph{Fluid Dyn. Res.}
       \textbf{27}, 367--404.
   \bibitem[Tailleux (2009)]{Tailleux2009}
      \textsc{Tailleux, R.} 2009
       On the energetics of stratified turbulent mixing, irreversible 
       thermodynamics, Boussinesq models, and the ocean heat
       engine controversy. \emph{J. Fluid Mech.}, in press.
   \bibitem[Weinstock (1978)]{Weinstock1978}
       \textsc{Weinstock, J.} 1978
        Vertical turbulent diffusion in a stably stratified fluid.
       \emph{J. Atmos. Sci.}, \textbf{62}, 3177--3180.
   \bibitem[Winters \& al. (1995)]{Winters1995}
       \textsc{Winters, K. B., Lombard, P. N., and Riley, J. J., 
       \& d'Asaro, E. A. (1995)}
       Available potential energy and mixing in density-stratified fluids.
       \emph{J. Fluid Mech.} \textbf{289}, 115--228.
    \bibitem[Wunsch \& Ferrari (2004)]{Wunsch2004}
        \textsc{Wunsch, C. \& Ferrari, R.}{2004}
        Vertical mixing, energy, and the general circulation of the oceans.
        \emph{Ann. Rev. of Fluid Mech.}, \textbf{36}, 281--314.

...

\end{thebibliography}
\end{document}